\documentclass[aps,prb,reprint,groupedaddress,preprintnumber,showpacs]{revtex4-1}
\usepackage{graphicx}
\usepackage{color}
\usepackage{dcolumn}

\begin{document}


\title{Ca$_3$Ir$_4$Sn$_{13}$: A weakly correlated nodeless superconductor}
\author{Kefeng Wang}
\affiliation{Condensed Matter Physics and Materials Science Department, Brookhaven National Laboratory, Upton, New York 11973, USA}
\author{C. Petrovic}
\affiliation{Condensed Matter Physics and Materials Science Department, Brookhaven National Laboratory, Upton, New York 11973, USA}

\date{\today}

\begin{abstract}
We report detailed Seebeck coefficient, Hall resistivity as well as specific heat measurement on Ca$_3$Ir$_4$Sn$_{13}$ single crystals. The Seebeck coefficient exhibits a peak corresponding to the anomaly in resistivity at $T^*$, and the carrier density is suppressed significantly below $T^*$. This indicates a significant Fermi surface reconstruction and the opening of the charge density wave gap at the supperlattice transition. The magnetic field induced enhancement of the residual specific heat coefficient $\gamma(H)$ exhibits a nearly linear dependence on magnetic field, indicating a nodeless gap. In the temperature range close to $T_c$ the Seebeck coefficient can be described well by the diffusion model. The zero-temperature extrapolated thermoelectric power is very small, implying large normalized Fermi temperature. Consequently the ratio $\frac{T_c}{T_F}$ is very small. Our results indicate that Ca$_3$Ir$_4$Sn$_{13}$ is a weakly correlated nodeless superconductor.
\end{abstract}

\maketitle

\section{Introduction}

Ca$_3$Ir$_4$Sn$_{13}$, a prominent member of Remeika phases, was found to exhibit superconducting transition with $T_c \sim 7$ K nearly thirty years ago.\cite{cairsn,cairsn1} It regains recent attention due to the possible coexistence of superconductivity and ferromagnetic spin-fluctuation as well as the three-dimensional charge density wave (CDW) from the supperlattice transition.\cite{cairsn2,cairsn3,cairsn4} The interaction between superconductivity, spin and charge fluctuations has been a central issue in the unconventional superconductivity.\cite{spin1,spin} In conventional phonon-mediated superconductors spin-fluctuations usually have negative effect on superconductivity, whereas they could be the source of the electron pairing in the vicinity of the quantum critical point (QCP) where spin-fluctuation is enhanced significantly in unconventional superconductors.\cite{spin1,spin2,spin3} It is expected that the spin-fluctuation pairing mechanism results in an unconventional superconducting states. Examples include $d$-wave gap in cuprates and heavy fermions superconductor CeCoIn$_5$, and the $p$-wave gap in Sr$_2$RuO$_4$.\cite{gap1,gap2,gap3}

The peak-like anomalies in resistivity and susceptibility of Ca$_3$Ir$_4$Sn$_{13}$ were attributed to the ferromagnetic spin-fluctuation, and the thermodynamic measurements suggested a strongly correlated system. Moreover, the resistivity shows a non-Fermi-liquid behavior in the normal state. Fermi liquid emerges by the suppression of the spin-fluctuation with increasing magnetic field.\cite{cairsn2} However, single crystal x-ray diffraction shows that the resistivity anomaly most likely corresponds to a temperature driven structural transition at $T^*$ from a simple cubic phase to the supperlattice with the doubled lattice parameters. This is related to the Fermi surface nesting along the body diagonal direction and the three dimensional charge density wave (CDW) instability of the conduction electron system.\cite{cairsn3}

In this paper, we report detailed Seebeck coefficient, Hall resistivity as well as specific heat measurement on Ca$_3$Ir$_4$Sn$_{13}$ single crystals.  The Seebeck coefficient exhibits a peak corresponding to the anomaly in resistivity at $T^*$, and the carrier density is suppressed significantly below $T^*$. This indicates a significant Fermi surface reconstruction and the opening of the charge density wave gap at the supperlattice transition. The magnetic field induced enhancement of the residual specific heat coefficient $\gamma(H)$ exhibits nearly linear dependence on magnetic field, indicating a nodeless gap. In the temperature range close to $T_c$ the Seebeck coefficient can be described well by the diffusion model. The zero-temperature extrapolated thermoelectric power is very small, implying large normalized Fermi temperature. Consequently the ratio $\frac{T_c}{T_F}$ is very small. Our results indicate that Ca$_3$Ir$_4$Sn$_{13}$ is a weakly correlated nodeless superconductor with no significant contribution of spin fluctuations to electronic system in the normal state.

\section{Experimental}

Single crystals of Ca$_3$Ir$_4$Sn$_{13}$ were grown using a high-temperature self-flux method.\cite{cairsn,cairsn1} X-ray diffraction (XRD) data were taken with Cu K$_{\alpha}$ ($\lambda=0.15418$ nm) radiation of Rigaku Miniflex powder diffractometer and the elemental analysis was performed using an energy-dispersive x-ray spectroscopy (EDX) in a JEOL JSM-6500 scanning electron microscopy. Electrical transport measurements were conducted in Quantum Design PPMS-9 with conventional four-wire method and the contacts were made directly to the crystal surface using silver epoxy. Thermal transport properties were measured in Quantum Design PPMS-9 from 2 K to 350 K using one-heater-two-thermometer method. In electric and thermal transport measurements, the magnetic field is always perpendicular to the heat/electrical current. The relative error in our measurement was $\frac{\Delta \kappa}{\kappa}\sim$5$\%$ and $\frac{\Delta S}{S}\sim$5$\%$ based on Ni standard measured under identical conditions. The specific heat measurements were measured in PPMS-9 using relaxation method under ambient pressure. Magnetic measurement were performed in a Quantum Design MPMS-5.

\section{Results and discussions}

Fig. 1(a) shows the powder XRD pattern at the room temperature of flux grown Ca$_3$Ir$_4$Sn$_{13}$ crystals, which was fitted by RIETICA software.\cite{rietica} All reflections can be indexed in the $Pm\bar{3}n$ space group (inset of Fig. 1(b)) and the refined lattice parameters are $a=b=c=9.708(6)~{\AA}$, which confirm the high quality single crystal and the absence of impurities of our samples.\cite{cairsn} Fig. 1(b) shows the magnetic susceptibility $4\pi\chi$ for single crystal below 20 K. The sharp drop in $4\pi\chi$ around 7 K demonstrates the superconducting transition which is consistent with previous report.\cite{cairsn,cairsn2} The superconducting volume ratio approaches $90\%$ which also confirms the high quality of our crystals.

\begin{figure}[tbp]
\includegraphics[scale=0.38]{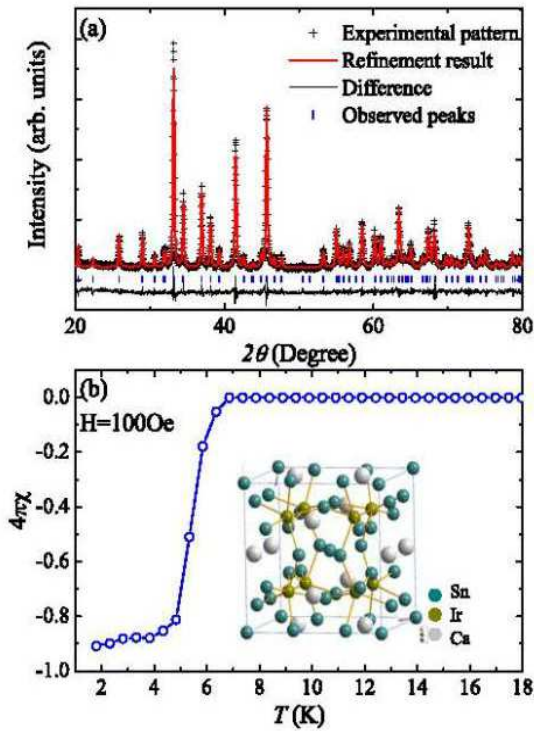}
\caption{(Color online) (a) Powder XRD patterns and structural refinement results. (b) Magnetic susceptibility $4\pi\chi$ for single crystal around superconducting transition temperature $T_c=7$ K in 100 Oe field. Inset in (b) shows the crystal crystal structure of Ca$_3$Ir$_4$Sn$_{13}$.}
\end{figure}

Temperature dependence of the electrical resistivity $\rho(T)$, Seebeck coefficient $S(T)$ and magnetization $M(T)$ in whole temperature range and different magnetic fields for Ca$_3$Ir$_4$Sn$_{13}$ single crystal is shown in Fig. 2. The resistivity $\rho(T)$ (Fig. 2(a)) is metallic behavior above $\sim 7$ K and exhibits a distinct anomaly at $T^*\sim$ 35 K (as shown by the dashed blue line in Fig. 2). It enters superconducting state at $T_c\sim$ 7.5 K (the dashed green line in Fig. 2). The value of $T_c$ and  $T^*$ is consistent with our susceptibility measurement and previous reports.\cite{cairsn2} The Seebeck coefficient $S(T)$ (Fig. 2(b)) is positive in the whole temperature range. With decreasing temperature, Seebeck coefficient decreases and shows a significant peak at $T^*$. $S(T)$ vanishes below $T_c$ since Cooper pairs carry no entropy.\cite{superconductor} 9 T external field totally suppresses the superconductivity and resisitivity/Seebeck coefficient becomes nonzero.

\begin{figure}[tbp]
\includegraphics[scale=1]{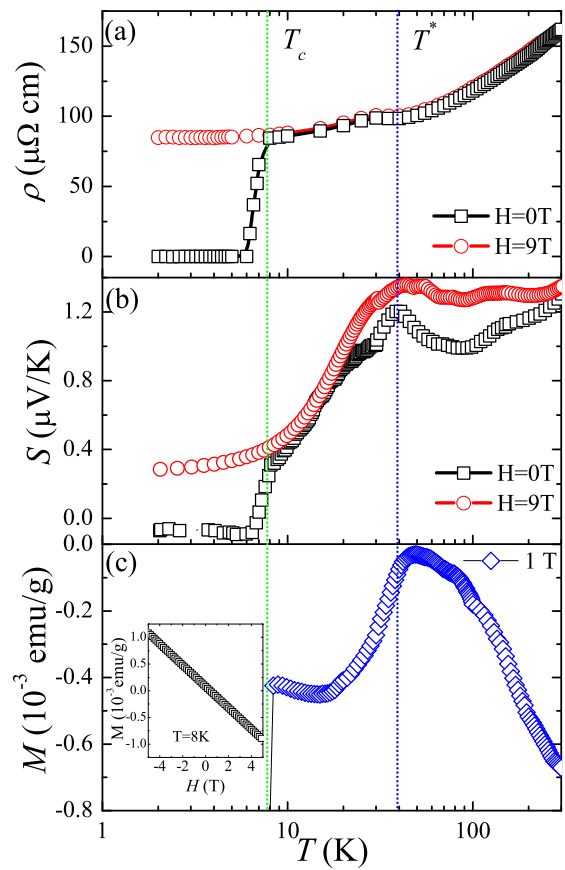}
\caption{(Color online) (a) Temperature dependence of the resistivity $\rho (T)$ (a), Seebeck coefficient $S(T)$ (b) and magnetization $M(T)$ (c) for Ca$_3$Ir$_4$Sn$_{13}$ single crystal in different magnetic fields respectively. Insets in (c) shows the magnetization-magnetic field curve at 8 K.}
\end{figure}

The similar anomaly in $\rho$ and $S$ at $T^*$ is also observed in magnetization (Fig. 2(c)). However, in contrary to the previous report where the magnetization measurement shows weak magnetic signal and spin fluctuation,\cite{cairsn2} our sample shows weak diamagnetic behavior in whole temperature range except in superconducting region (Fig. 2(c)). The magnetization-field loop (inset in Fig. 2(c)) up to 5 T field at 8 K confirms the diamagnetic state. Our results are consistent with the magnetic measurements in Sr$_3$Ir$_4$Sn$_{13}$ and argue against the signatures of spin fluctuations.\cite{cairsn3}

\begin{figure}[tbp]
\includegraphics[scale=1]{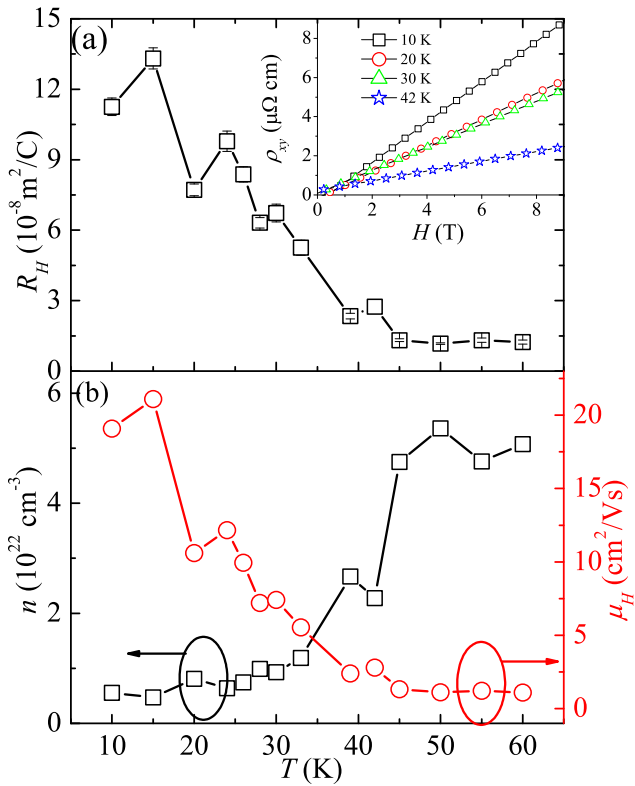}
\caption{(Color online) Temperature dependence of the the Hall coefficient $R_H$ (a), the carrier density $n$ and mobility $\mu_H$ for Ca$_3$Ir$_4$Sn$_{13}$ crystal.  Inset in (a) shows the Hall resistivity $\rho_{xy}$ at four different temperatures.}
\end{figure}

Fig. 3(a) shows the temperature dependence of the Hall coefficient $R_H$. $R_H$ is positive and consistent with the positive Seebeck coefficient, indicating hole-type carriers. With decreasing temperature, $R_H$ and correspondingly the carrier density $n=\frac{1}{e|R_H|}$ are nearly constant above $\sim$ 40 K (Fig. 3(b)). But $R_H$ increases significantly and consequently the carrier density is strongly suppressed below $T^*$. The Hall mobility $\mu_H=R_H/\rho$ also shows a sharp increase at $T^*$ with decreasing temperature (Fig. 3(b)).

The resistivity anomaly at $T^*\sim35$ K of Ca$_3$Ir$_4$Sn$_{13}$ was considered be from the ferromagnetic spin fluctuation since a negative magnetoresistance is observed which could be attributed to the suppression of the spin fluctuation by magnetic fields. The resistivity in the normal state shows non-Fermi-liquid behavior possibly due to the scattering related to the same spin fluctuation mechanism.\cite{cairsn2} However, the detailed single crystal XRD shows that (Sr/Ca)$_3$Ir$_4$Sn$_{13}$ undergoes a temperature driven structural transition at $T^*$ from a simple cubic phase to the supperlattice variant which has a doubled lattice parameter when compared to the high temperature phase. This transition is possibly related to the CDW instability of the conduction electron system due to the Fermi surface nesting along the body diagonal.\cite{cairsn3}

The CDW instability will induce the Fermi surface reconstruction and open the CDW gap. This always induces significant anomaly in Seebeck coefficient $S$ since it is directly related to the derivative of the density of states at Fermi level. For example, TiSe$_2$ at the CDW transition,\cite{ZrTe}, cuprates (such as YBa$_2$Cu$_3$O$_{6.67}$) \cite{cuprate1,cuprate2} and iron-based superconductors (such as BaFe$_2$As$_2$ and SmFeAsO$_{0.85}$)\cite{iron1,iron2,iron3} at the spin density wave transition, exhibit distinct anomaly in Seebeck coefficient attributed to the corresponding Fermi surface reconstruction. The anomaly in Seebeck coefficient at $T^*$ in Ca$_3$Ir$_4$Sn$_{13}$ reflects the Fermi surface reconstruction and the suppression of carrier density below $T^*$.

Alternative explanation for the peak in $S$ is the spin entropy due to the spin fluctuation such as the Seebeck coefficient in Na$_x$CoO$_2$.\cite{NaCoO2} However, our crystal shows diamagnetic behavior in whole temperature range. Besides, the magnetic field has no significant influence on the Seebeck coefficient in low temperature range (except for temperature range below $T_c$ where the magnetic fields destroy the superconductivity and enhance the Seebeck coefficient as shown in Fig. 2(b), and also does not change the peak position in Seebeck coefficient. Our measurements reveals a small increase in Seebeck coefficient in magnetic field and high temperature range (around $10\%$). This is also opposite to the negative magnetothermopower effect observed in Na$_x$CoO$_2$ since the magnetic field will suppress the spin fluctuation.\cite{NaCoO2} These results rule out the spin fluctuation in our crystals. Besides these, the suppression of carrier density shown by Hall resistivity measurement also implies the opening of the gap at $T^*$. Base on these discussion, the peak in Seebeck coefficient, resistivity and magnetization should originate from the CDW transition at $T^*$.

\begin{figure}[tbp]
\includegraphics[scale=0.95]{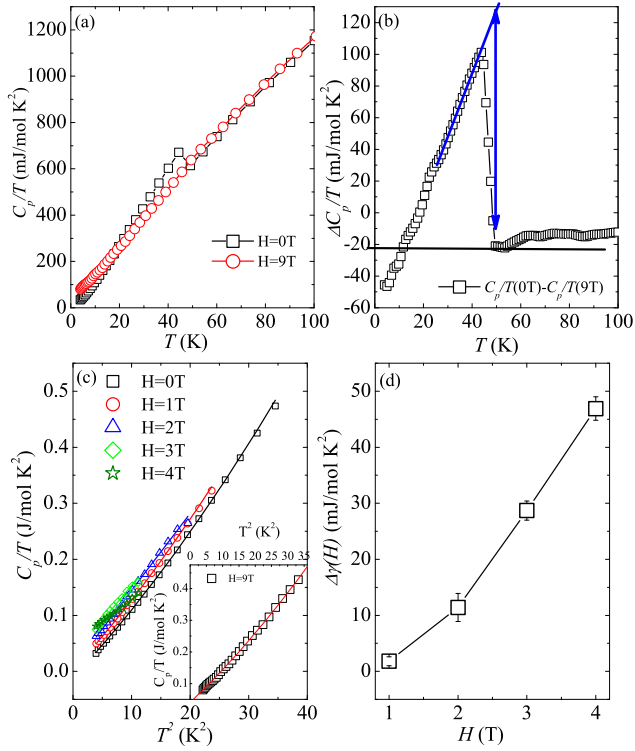}
\caption{(Color online) (a) Temperature dependence of the specific heat $C_p$ of Ca$_3$Ir$_4$Sn$_{13}$ single crystal in 0 T (squares) and 9 T (circles) respectively. (b) The difference in the specific heat data between 0 T and 9 T, $\Delta C_p/T = C_p(H)-C_p(0)$. The blue solid lines here are just to show determining the height of the specific heat anomaly. (c) Low temperature specific heat $C_p/T$ vs $T^2$ in different magnetic fields up to 4 T Ca$_3$Ir$_4$Sn$_{13}$ single crystal . The lines are the fitting results below the superconducting transition temperature $T_c$ using $C_p(T,H)=\gamma(H)T+\beta T^2+\eta T^5$. (d) Field dependence of the Sommerfeld coefficient $\Delta\gamma (H)=\gamma(H)-\gamma(0)$ for crystal.}
\end{figure}

Fig. 4(a) shows the heat capacity of Ca3Ir4Sn14 near superconducting transition whereas Fig. 4(b) shows the difference between the specific heat between 0 T and 9 T. Corresponding to the superconducting transition, the heat capacity $C_p(T)$ shows a jump at $T_c$, but there is no heat capacity anomaly at $T^*$. Superconducting transition is suppressed in 9 T magnetic field. The height of the specific heat anomaly $\Delta C_p/T|_{T_c}$ near $T_c$ is estimated to be $124\pm1$ mJ/mol K$^2$. The specific heat data for the sample below superconducting transition temperature in different magnetic fields are plotted as $C_p/T$ vs. $T^2$ in Fig. 4(c). All data could be fitted well by $C_p(T,H)=\gamma(H)T+\beta T^3+\eta T^5$, where $\gamma(H)$ is the residual specific heat coefficient in the magnetic field and $\beta T^3+\eta T^5$ is the phonon contribution. In the fitting it was necessary to include anharmonic phonon term $\sim T^5$, possibly due to the very small Debye temperature (fitting of the specific data in 9 T field where superconductivity is completely suppressed gives $\Theta_D\sim150$ K, as shown in the inset of Fig. 4(c)).\cite{hc1} Magnetic field enhances the residual specific heat coefficient $\gamma$ progressively due to the generation of the quasiparticle density of states (Fig. 4(c)). The magnetic field induced enhancement of $\gamma(H)$, $\Delta\gamma(H)=\gamma(H)-\gamma(0)$, obtained from the fitting procedure above, is shown in Fig. 4(d). In superconductors with nodal gap, such as cuprates, $\Delta\gamma(H)$ exhibits a square root field dependence, i.e. $\Delta\gamma(H)\propto\sqrt{H}$, and the residual heat capacity term is large.\cite{hc2,hc3,hc4,hc5,hc6} However, in Ca$_3$Ir$_4$Sn$_{13}$, $\Delta\gamma(H)$ shows nearly linear field dependence and the residual heat capacity term is very small. Linear dependence of heat capacity term was also observed in nodeless superconductors such as K$_x$Fe$_{2-y}$Se$_2$,\cite{hc1} and points to the nodeless gap in Ca$_3$Ir$_4$Sn$_{13}$.

Another important aspect of superconductivity in Ca$_3$Ir$_4$Sn$_{13}$ is the proper characterization of the electronic correlation strength in the normal state. Seebeck coefficient was used to characterize the correlation strength in several superconductors, such as FeTe$_{1-x}$Se$_x$, cuprates and LuNi$_2$B$_2$C.\cite{TEP1,TEP2,TEP3,TEP4,TEP5,TEP6} We now turn to Seebeck coefficient in the normal state. Diffusive Seebeck response of a Fermi liquid is expected to be linear in $T$ in the zero-temperature limit, with a magnitude proportional to the strength of electronic correlations. This is similar to the T-linear electronic specific heat, $C_e/T=\gamma$. Both can be linked to the Fermi temperature $T_F=\epsilon_F/k_B$:
\begin{eqnarray}
S/T & = & \pm\frac{\pi^{2}}{2}\frac{k_B}{e}\frac{1}{T_F} \\
\gamma & = & \frac{\pi^2}{3}k_B\frac{n}{T_F}
\end{eqnarray}
where $k_B$ is Boltzmann's constant, $e$ is the electron charge, $\epsilon_F$ is the Fermi energy and $n$ is the carrier density.\cite{TEP7,TE} Fig. 5 presents the temperature dependence of Seebeck coefficient divided by $T$, $S/T$ under 0 T for Ca$_3$Ir$_4$Sn$_{13}$ crystal. $S/T$ in the normal state near $T_c$ is nearly linear and can be described well by diffusive model. The zero-temperature extrapolated value of $S/T$ is $\sim0.034(7)$ $\mu V/K$. We can therefore extract T$_{F}$ = 12,500 K.

\begin{figure}[tbp]
\includegraphics[scale=0.32]{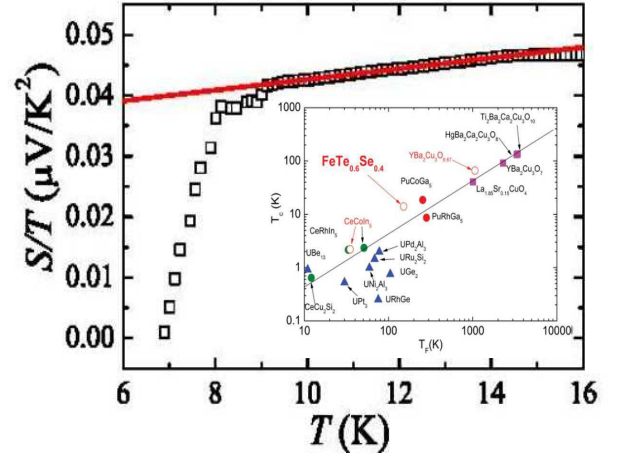}
\caption{(Color online) The temperature dependence of Seebeck coefficient divided by $T$, $S/T$ under 0 T for Ca$_3$Ir$_4$Sn$_{13}$ crystal. The open squares are the experimental data and the solid line is the linear fitting result within normal phase. Inset: $T_c$ as a function of Fermi temperature $T_F$  in a few superconductors. The black circle and arrow shows the position of Ca$_3$Ir$_4$Sn$_{13}$. Adapted from Ref.[33,35].}
\end{figure}

The ratio of the superconducting transition temperature to the normalized Fermi temperature $\frac{T_c}{T_F}$ characterizes the correlation strength in superconductors.\cite{TEP3,TEP5,TEP7,TEP8} In unconventional superconductors, such as CeCoIn$_5$, YBa$_2$Cu$_3$O$_{6.67}$ and iron-based superconductors,\cite{TEP3,TEP5,TEP7,TEP8} this ratio is about 0.1 (the inset in Fig. 5), but it is only $\sim0.02$ in BCS superconductors (such as LuNi$_2$B$_2$C) and only around 0.005 in Nd.\cite{TEP5,TEP7,TEP8} In Ca$_3$Ir$_4$Sn$_{13}$ crystal, this ratio $\frac{T_c}{T_F}$ is only around 0.001 (as shown by the red arrow and filled circle in the inset of Fig. 5). This implies that electronic correlations in Ca3Ir4Sn13 are weak. This is in contrast to what is commonly found in unconventional superconductors near magnetic instability. This gives an argument against unconventional pairing mechanism associated with spin fluctuations.\cite{cairsn2,spin2}

The fitting of the specific heat data in 9 T field gives the linear electron specific heat coefficient in normal state $\gamma_n=39\pm3$ mJ/mol K$^2$ (inset of Fig. 3(a)), which is close to the previous report.\cite{cairsn2} The absolute value of the dimensionless ratio of Seebeck coefficient $S/T$ to specific heat term $\gamma_n$, $q=\frac{N_{Av}eS}{T\gamma}$, with $N_{Av}$ the Avogadro number, provides the carrier density. Calculation gives the carrier density with $|q|^{-1}\simeq 11.8(8)$ carrier per unit cell, which is consistent with the large Fermi surface by first principle calculations.\cite{cairsn3} Taking into account that a superlattice transition happens and the a lattice parameter is doubled below $T^*$, the calculated carrier density $n=|q|^{-1}/\Omega\sim 6\times10^{21}$ cm$^{-3}$ (where $\Omega$ is the unit cell volume), which is consistent with the Hall carrier density in Fig. 3(b).

\section{Conclusion}

We report detailed thermal, transport and thermodynamic characterization of Ca$_3$Ir$_4$Sn$_{13}$ single crystals. The Seebeck coefficient exhibits an peak corresponding to the anomaly in resistivity at $T^*$, and the carrier density is suppressed significantly below $T^*$. This indicates a significant Fermi surface reconstruction and the opening of the charge density wave gap at the supperlattice transition. The magnetic field induced enhancement of the residual specific heat coefficient $\gamma(H)$ exhibits nearly linear dependence on magnetic field, indicating a nodeless gap. In the temperature range close to $T_c$ the Seebeck coefficient can be described well by diffusion model. The zero-temperature extrapolated thermoelectric power is very small, implying large normalized Fermi temperature. Consequently the ratio $\frac{T_c}{T_F}$ is very small. Our results indicate that Ca$_3$Ir$_4$Sn$_{13}$ is a weakly correlated nodeless superconductor.

\begin{acknowledgments}
We than John Warren for help with SEM measurements. Work at Brookhaven is supported by the U.S. DOE under contract No. DE-AC02-98CH10886.
\end{acknowledgments}


\end{document}